\documentclass[aps,prl,reprint,superscriptaddress]{revtex4-2}

\usepackage{graphicx}
\usepackage{dcolumn}
\usepackage{bm}
\usepackage{hyperref}

\usepackage{color}

\begin{document}

\preprint{APS/123-QED}

\title{Broken Kramers{\textquoteright} Degeneracy in Altermagnetic MnTe}

\author{Suyoung Lee}
 \affiliation{Center for Correlated Electron Systems, Institute for Basic Science, Seoul 08826, Korea}
 \affiliation{Department of Physics and Astronomy, Seoul National University, Seoul 08826, Korea}

 \author{Sangjae Lee}
 \affiliation{The Research Institute of Basic Sciences, Seoul National University, Seoul 08826, Korea}

 \author{Saegyeol Jung}
 \affiliation{Center for Correlated Electron Systems, Institute for Basic Science, Seoul 08826, Korea}
 \affiliation{Department of Physics and Astronomy, Seoul National University, Seoul 08826, Korea}

 \author{Jiwon Jung}
 \affiliation{Center for Artificial Low Dimensional Electronic Systems, Institute for Basic Science (IBS), Pohang 37673, Korea}
 \affiliation{Department of Physics, Pohang University of Science and Technology (POSTECH), Pohang 37673, Republic of Korea}

 \author{Donghan Kim}
 \affiliation{Center for Correlated Electron Systems, Institute for Basic Science, Seoul 08826, Korea}
 \affiliation{Department of Physics and Astronomy, Seoul National University, Seoul 08826, Korea}

 \author{Yeonjae Lee}
 \affiliation{Center for Correlated Electron Systems, Institute for Basic Science, Seoul 08826, Korea}
 \affiliation{Department of Physics and Astronomy, Seoul National University, Seoul 08826, Korea}

 \author{Byeongjun Seok}
 \affiliation{Center for Correlated Electron Systems, Institute for Basic Science, Seoul 08826, Korea}
 \affiliation{Department of Physics and Astronomy, Seoul National University, Seoul 08826, Korea}

 \author{Jaeyoung Kim}
 \affiliation{Center for Artificial Low Dimensional Electronic Systems, Institute for Basic Science (IBS), Pohang 37673, Korea}

 \author{Byeong Gyu Park}
 \affiliation{Pohang Accelerator Laboratory, Pohang University of Science and Technology, Pohang 37673, Korea}

  \author{Libor \v Smejkal}
  \affiliation{Institut f\"ur Physik, Johannes Gutenberg Universit\"at Mainz, D-55099 Mainz, Germany}
  \affiliation{Institute of Physics, Czech Academy of Sciences, Cukrovarnick\'a 10, 162 00 Praha 6 Czech Republic}

  \author{Chang-Jong Kang}
 \email{cjkang87@cnu.ac.kr}
 \affiliation{Department of Physics, Chungnam National University, Daejeon 34134, Korea}

 \author{Changyoung Kim}
 \email{changyoung@snu.ac.kr}
 \affiliation{Center for Correlated Electron Systems, Institute for Basic Science, Seoul 08826, Korea}
 \affiliation{Department of Physics and Astronomy, Seoul National University, Seoul 08826, Korea}

\date{\today}

\begin{abstract}
Altermagnetism is a newly identified fundamental class of magnetism with vanishing net magnetization and time-reversal symmetry broken electronic structure.
Probing the unusual electronic structure with nonrelativistic spin splitting would be a direct experimental verification of altermagnetic phase.
By combining high-quality film growth and $in~situ$ angle-resolved photoemission spectroscopy, we report the electronic structure of an altermagnetic candidate, $\alpha$-MnTe.
Temperature dependent study reveals the lifting of Kramers{\textquoteright} degeneracy accompanied by a magnetic phase transition at $T_N=267\text{ K}$ with spin splitting of up to $370\text{ meV}$, providing direct spectroscopic evidence for altermagnetism in MnTe.

\end{abstract}

\maketitle

Altermagnetism has recently been theoretically identified to be the third ground state of collinear magnetism, characterized by alternating spin characters both in real and reciprocal spaces~\cite{Smejkal2022b,Mazin2022}. Although previously classified as antiferromagnets for their vanishing net magnetization, it has been shown that altermagnets are fundamentally distinct from antiferromagnets (AFMs) in the nonrelativistic spin group formalism~\cite{Brinkman1966,LITVIN1974538,Litvin1977}. The broken time-reversal symmetry leads to spin split electronic structure even without spin-orbit interactions~\cite{Smejkal2020,Nakatsuji2022}. Owing to its unique spin-momentum locked electronic structure, altermagnet has been proposed to exhibit distinct time-reversal symmetry breaking phenomena suitable for various spintronics applications~\cite{Smejkal2022e}.

Despite having been vigorously explored in theory~\cite{Ahn2019,Mazin2021,Smejkal20223,Cui2023prb,Steward2023,Ghorashi2023,Cui2023,Zhang2023,Ouassou2023,Zhu2023t,Papaj2023,Zhou2023,Hariki2023}, direct verification of altermagnetic state via spectroscopic methods $e.g.$ measuring the spin-split bands by angle-resolved photoemission spectroscopy (ARPES) remains challenging. First of all, compounds with giant spin splitting are relatively rare so far although altermagnetic phases are found in a wide range of materials~\cite{Smejkal2022e}. More importantly, spin polarization in the experimentally measured band structure may cancel out due to contributions from mixed domains as detwinning altermagnetic domains is generally nontrivial. As a result, direct spectroscopic evidence for spin-split electronic bands with lifted Kramers{\textquoteright} degeneracy remains elusive despite recent report on unusual circular dichroism~\cite{Fedchenko2023} texture of an altermagnet in photoemission.

\begin{figure*}
\includegraphics{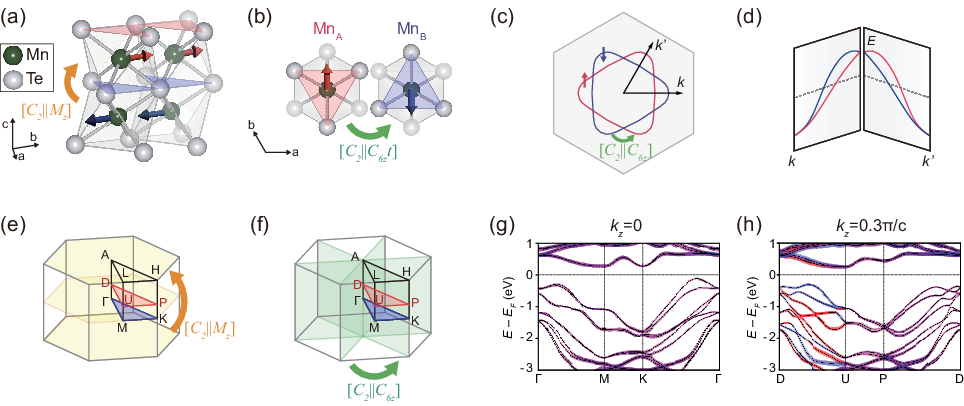}
\caption{\label{fig1} (a),(b)Crystal structure of MnTe with two opposite-spin sublattices ($\text{Mn}_{\text{A}}$ and $\text{Mn}_{\text{B}}$) and possible sublattice-transposing transformations containing a real-space mirror or sixfold rotation.
(c) Schematic constant-energy contour on a $k_{x}-k_{y}$ plane with $k_{z}\neq n\pi/c$ corresponding to a spin group $[C_{2}{\parallel}C_{6z}]$. 
(d) Schematic band dispersion plotted along two momenta $k$ and $k'$ marked in (c), showing spin-split electronic structure with alternating spin polarizations.
(e),(f) 3D Brillouin zone and spin-degenerate nodal planes protected by (e) $[C_{2}{\parallel}M_{z}]$ and (f) $[C_{2}{\parallel}C_{6z}]$. (g),(h) DFT calculated spin-resolved electronic structures of MnTe plotted on (g) nodal ($k_{z}=0$) and (h) off-nodal ($k_{z}=0.3\pi/c$) plane.
Here, the red and blue colors correspond to spin-up and down, respectively.
The spin splitting is clearly realized along the off-nodal D-U line.
}
\end{figure*}

$\alpha$-MnTe is a suitable material for electronic structure study on altermagnets. As illustrated in Fig.~\ref{fig1}(a) and (b), MnTe has a hexagonal NiAs-type structure (space group $P6_{3}/mmc$)~\cite{Villars2023:sm_isp_sd_0379437}. Below the N\'eel temperature ($T_{N}=307\text{ K}$ for bulk), it exhibits an A-type AFM ground state where the spins align ferromagnetically along the $[1\overline{1}00]$ axis on the $ab$-plane and antiferromagnetically between adjacent layers~\cite{Kriegner2017}.  The opposite-spin sublattices, Mn$_{\text{A}}$ and Mn$_{\text{B}}$, are not connected by a simple space inversion or translation but instead by a sixfold rotation or a mirror operation, which leads to alternating spin polarizations in the electronic structure as schematically shown in Fig.~\ref{fig1}(c) and (d). MnTe has been predicted to host one of the largest spin splittings among those of known altermagnetic candidates~\cite{Smejkal2022b}. Previous study on the transport properties of MnTe has reported observation of spontaneous anomalous Hall effect arising from altermagnetic order~\cite{Gonzalez2023}. However, ARPES measurement on single crystalline MnTe is challenging due to the 3D crystal structure.

Growing epitaxial films is an alternative way to achieve a clean surface of a 3D crystal, enabling surface-sensitive measurements such as ARPES. In this Letter, we report successful growth of high quality MnTe films by molecular beam epitaxy (MBE) and $in\:situ$ ARPES studies on the films. Our temperature-dependent ARPES data firmly establish breaking of Kramers{\textquoteright} degeneracy accompanied by the magnetic transition in MnTe, which is consistent with density functional theory (DFT) calculation results. This work not only provides the first spectroscopic evidence for spin-split electronic structure of altermagnets but also helps understand and further engineer intriguing spin-dependent phenomena in MnTe, including the previously-reported anomalous Hall effect.

Collinear magnetic phases can be classified based on nonrelativistic spin groups
that involve transformations $[R_{i}{\parallel}R_{j}]$
where $R_i$ operates solely in spin space, while $R_j$ acts in real space~\cite{LITVIN1974538,Litvin1977}.
Altermagnetism is characterized by the interplay of two rotations:
a twofold rotation in spin space combined with a rotation in real space,
denoted as $[C_{2}{\parallel}A$] ($A$: a rotational operator).
This distinct symmetry connects the sublattices with opposite spins through rotation, rather than via space inversion or translation.
As a consequence, the effective time-reversal symmetry is broken
in the electronic structure of an altermagnet~\cite{Smejkal2022b,Smejkal2022e}.

In contrast to the spin-momentum-locked phases arising from broken space-inversion symmetry and relativistic spin-orbit coupling (SOC),
nonrelativistic spin splitting in an altermagnet is not universally prohibited at time-reversal invariant momenta~\cite{Smejkal20223}.
Instead, spin-degenerate nodes are protected by the symmetry of the spin group.
For instance, both space-inversion and $[C_{2}{\parallel}A$] symmetries preserve
spin degeneracy at the $\Gamma$ point~\cite{Smejkal2022b}.
In the case of MnTe, the $[C_{2}{\parallel}M_{z}]$ symmetry,
encompassing a real-space mirror operation with respect to the $ab$-plane,
establishes spin-degenerate nodal planes at $k_{z}=n\pi/c$
(where $n$=integer, $c$=out-of-plane lattice parameter),
which is highlighted in Fig.~\ref{fig1}(e).
Similarly, the $[C_{2}{\parallel}C_{6z}t]$ symmetry
upholds spin degeneracy across all $\Gamma$KHA planes (Fig.~\ref{fig1}(f)).
It is noteworthy that all high-symmetry lines within a hexagonal Brillouin zone pertain to such nodal planes.
Thus, it is necessary to investigate off-symmetric regions in the Brillouin zone,
such as a constant-$k_{z}$ plane with $k_{z}\neq n\pi/c$,
in order to comprehensively capture the spin-split electronic structure of MnTe.

First, we theoretically confirm the existence of spin-split electronic structure of altermagnetic MnTe using DFT. Details regarding the methodology of theoretical calculations are illustrated in the Supplemental Materials (SM) Sec. I~\cite{SM}.
Fig.~\ref{fig1}(g) shows the calculated electronic structure and spin character along high-symmetry lines $\Gamma$-M-K-$\Gamma$. Although the inclusion of SOC induces additional band splittings along nodal planes~\cite{Yin2019}, no net spin polarization is allowed due to symmetry constraints. On the other hand, spin-split bands are observed along an off-nodal line D-U, which is a line parallel to $\Gamma$-M, but with $k_{z}=0.3\pi/c$ (Fig.~\ref{fig1}(h)).

\begin{figure}
\includegraphics{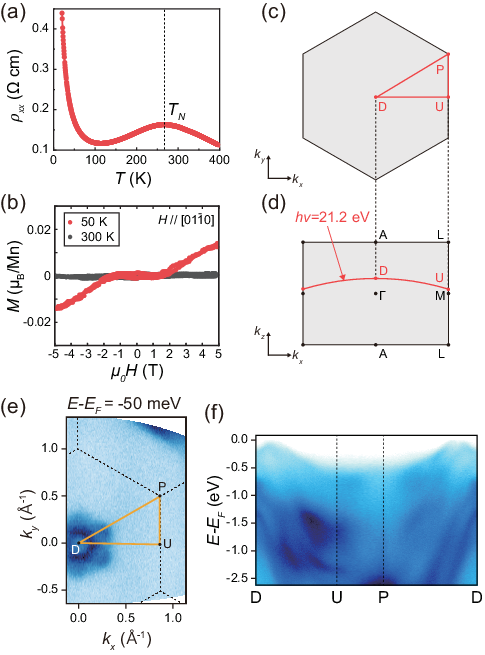}
\caption{\label{fig2} Physical properties of MnTe films.
(a) Temperature dependent longitudinal resistivity $\rho_{xx}$ for $T$ between $20$ and $400\text{ K}$. (b) Magnetization for a magnetic field direction $H//[01\overline{1}0]$ measured at $T=50\text{ K}$ and $300\text{ K}$, showing absence of net magnetization at $H=0$. Diamagnetic contribution from the substrate is subtracted.
(c),(d) Surface Brillouin zones of $k_{x}-k_{y}$(a) and $k_{x}-k_{z}$(b) planes. The curve highlighted in (d) corresponds to the momentum space where ARPES is measured with photon energy $h\nu=21.2\text{ eV}$.
(e) Constant energy ARPES map taken at $E-E_{F}=-50\text{ meV}$. Dotted lines mark the surface Brillouin zone.
(f) ARPES $E-k$ cut along the high symmetry lines in the surface Brillouin zone highlighted in (e).}
\end{figure}

MBE-grown MnTe film was adopted as a platform to experimentally study the unique altermagnetic ground state. Basic characterizations of the film can be found in Supplemental Material Sec. II~\cite{SM}. The temperature dependent longitudinal resistivity of a 50 nm-thick MnTe is shown in Fig.~\ref{fig2}(a). The N\'eel temperature can be determined from the peak in the resistivity curve~\cite{Magnin2012} to be $T_{N}=267\text{ K}$, which is slightly lower but consistent with the bulk value~\cite{Komatsubara1963}. The decrease in resistivity from $T=250\text{ K}$ to $50\text{ K}$ has been attributed to interactions between itinerant and local spins~\cite{Magnin2012}. The magnetization curve (Fig.~\ref{fig2}(b)) showing no spontaneous net magnetization further confirms that the film is in the $\alpha$-MnTe phase.

As shown in the DFT calculation, MnTe has a 3D electronic structure with spin-degenerate nodal planes at $k_{z}=n\pi/c$. In ARPES, by use of a light source with a fixed photon energy one can selectively map a $k_{x}-k_{y}$ surface with a specific $k_{z}$ in the 3D Brillouin zone.
Photon-denergy dependent ARPES reveals that the photon energy of 21.2 eV He-I$\alpha $ corresponds to $k_{z}=0.3\pi/c$ as shown in Fig.~\ref{fig2}(d) (See Supplemental Material Sec.III~\cite{SM}). By mapping the bands onto the surface Brillouin zone as illustrated in Fig.~\ref{fig2}(c), we expect to observe spin-polarized bands as predicted in the calculated electronic structure in Fig.~\ref{fig1}(h).

The constant energy contour taken at $E-E_{F}=-50\text{ meV}$ using 21.2 eV photon is shown in Fig.~\ref{fig2}(e).
Although MnTe is a semiconductor with a gap of 1.27 eV~\cite{Ferrer2000,Mobasser1985,Kriegner2016}, the observed bands have small spectral weight at the Fermi energy due to natural $p$-doping by excess Mn, which has been previously reported in bulk~\cite{Ferrer2000} and thin film MnTe~\cite{Kriegner2016,Mori2018}. 
The observed bands show sixfold rotational symmetry with valence band maximum near the D point, which is consistent with the crystal symmetry of MnTe and the DFT calculation. The high-symmetry cut in Fig.~\ref{fig2}(f) shows good agreement with the DFT calculation presented in Fig.~\ref{fig1}(h).

\begin{figure}
\includegraphics{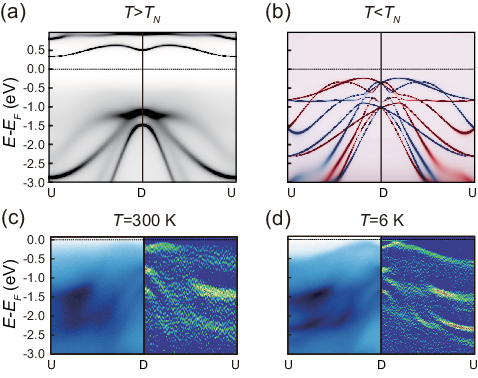}
\caption{\label{fig3}Temperature-dependent evolution of MnTe electronic structure.
DMFT results in the (a) paramagnetic ($T=2321\text{ K}$) and (b) altermagnetic ($T=116\text{ K}$) states.
Here, the red and blue lines correspond to spin-up and down states, respectively.
ARPES cut (left) and its second derivative (right) taken along the D-U direction in the (c) paramagnetic and (d) altermagnetic states taken at $T=300\text{ K}$ and $6\text{ K}$, respectively.
}
\end{figure}

To verify that Kramers{\textquoteright} degeneracy is lifted in the measured bands, we also performed DFT+dynamical mean field theory (DMFT) calculations and temperature-dependent ARPES to compare the electronic structures of MnTe in altermagnetic and paramagnetic states. Fig.~\ref{fig3}(a) is the DMFT results along the off-nodal D-U line for paramagnetic MnTe ($T>T_N$). Bands are spin degenerate at all momenta due to the recovered time reversal symmetry. In the altermagnetic state ($T<T_N$), the number of bands doubles as the initially doubly-degenerate bands split into spin-polarized bands except for symmetry-protected nodes (D and U). The antiparallel spin polarizations alternating in the Brillouin zone shown in Fig.~\ref{fig3}(b) are characteristic features of the altermagnetism.

ARPES spectra were obtained for both paramagnetic and altermagnetic MnTe along the D-U line
at $T = 300~\text{K}$ and $T = 6~\text{K}$, respectively.
Notably, the DMFT simulation at $T > T_{N}$ shows precise correspondence to the ARPES spectra at $T = 300~\text{K}$ with an energy offset of $\sim$0.23 eV (see Figs.~\ref{fig3}(a) and (c)).
This clear alignment strongly supports the validity of the current DMFT calculations~\cite{SM}.
Both the ARPES and DMFT data reveal a doubling of the number of bands during the altermagnetic transition,
indicating the lifting of the Kramers' degeneracy due to broken time-reversal symmetry.
The ARPES and DMFT results suggest substantial spin splittings of a few hundred meVs between spin-polarized bands along the D-U line.

\begin{figure}
\includegraphics{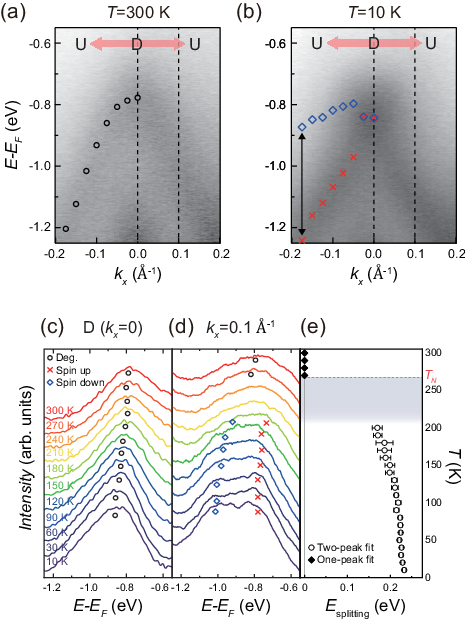}
\caption{\label{fig4}(a),(b) ARPES cuts taken along the D-U line at temperatures above (a) and below (b) the $T_N$.  
(c),(d) Detailed temperature-dependent evolution of the ARPES spectral function taken at (c) D and (d) $0.1\text{ \AA}^{-1}$ away from D. The fitted positions of the peaks are marked in black circle, red cross and blue diamond, denoting spin-degenerate, spin-up, and spin-down, respectively. The spin splitting at $k_{x}=-0.175~\text{\AA}^{-1}$ is marked in the black arrow.
(e) Temperature-dependent energy splitting between the spin-up and spin-down bands. Error bars are obtained from the standard errors of the fitted peak positions. Energy splittings in the shaded area near the $T_N$ ($210~{\leq}~T~{\leq}~260$ K) could not be reliably determined due to the uncertainty in the peak fitting at this temperature range.}
\end{figure}

To analyze the temperature dependence in more detail,
we focus on the hole band centered at D and
ascertain the peak positions through the application of Lorentzian fitting to the ARPES spectra.
Due to the D point being a spin-degenerate node protected by $[C_{2}{\parallel}C_{6z}]$,
only one peak is observed within the energy range of ($-1.2~\text{eV}$, $-0.6~\text{eV}$)
in both paramagnetic and altermagnetic states.
At $T = 300~\text{K}$, fitting the peak positions of the energy distribution curves (EDCs) from different momentum values
reveals a single parabolic hole band, as displayed in Fig.~\ref{fig4}(a).
At $T = 10~\text{K}$,
moving away from D towards U,
we find that the single peak in the EDC taken at D at $-0.9~\text{eV}$
splits into two peaks at off-nodal momenta, forming two hole bands crossing at D,
as marked in Fig.~\ref{fig4}(b). 
The splitting between the two bands is largest at $k_{x}=-0.175~\text{\AA}^{-1}$ with splitting energy of $370~\pm~30$ meV, which is consistent with the magnitude of the splitting found in the DMFT (Fig.~\ref{fig3}(b)).
Such momentum-dependent band splitting indicates that the two bands highlighted with markers in Fig.~\ref{fig4}(b) are opposite-spin pair bands originating from altermagnetism.

To further investigate the temperature-dependent evolution of the electronic structure,
we examine the EDCs at nodal and off-nodal momenta for various temperatures.
At D, as shown in Fig.~\ref{fig4}(c),
a single peak persists over all temperatures,
exhibiting only slight shifts in energy and thermal broadening in the spectra.
In contrast, at an off-nodal momentum, specifically at $k_{x}=0.1\text{ \AA}^{-1}$,
we observe that a single peak at $T = 300~\text{K}$ gradually splits into two spin-split peaks as the system is cooled down (Fig.~\ref{fig4}(d)).
The band splitting $E_{\text{splitting}}$ between the spin-split bands at $k_{x}=0.1\text{ \AA}^{-1}$ is plotted in Fig.~\ref{fig4}(e). 
$E_{\text{splitting}}$ at $T~{\leq}~200$ K clearly shows gradual decrease as the temperature increases, and vanishes above the $T_N$ ($T~\geq~270$ K). 

The reduced splitting and thermal broadening at raised temperatures hinder precise determination of the peak positions at temperature between $T=210$ to $260$ K (shaded area in Fig.~\ref{fig4}(e)).
Although exact fitting of $E_{\text{splitting}}$ to the mean field model $(1-T/T_{N})^{\beta}$ could not be made due to such uncertainty near the $T_N$, the temperature-dependent evolution of $E_{\text{splitting}}$ at $T{\ll}T_N$ and $T>T_N$ strongly suggests continuous lifting of Kramer{\textquoteright} degeneracy accompanied by second order magnetic phase transition.
Notably, the band presenting the energy splitting exhibits a significant dispersion with a bandwidth of $\sim$1.0 eV, indicative of an itinerant character.
Based on the ARPES and DMFT results,
we conclude that the energy splitting originates from spin separation along the off-nodal $k$-paths,
indicating a breaking of time-reversal symmetry along these paths.
This energy splitting in an altermagnet, to the best of our knowledge, is measured for the first time,
and it directly reflects the itinerant nature of altermagnetism.

In order to rule out the possibility of band splitting caused by structural transition inducing band folding or lifting of crystal centrosymmetry, we have performed temperature dependent low energy electron diffraction (LEED) and second harmonic generation (SHG). The results are presented in the Supplemental Material~\cite{SM}. We have not detected any signatures of structural reconstruction in the $ab$-plane near the $T_N$ in LEED. Furthermore, we employed the SHG technique which is highly sensitive to breaking of the space-inversion symmetry ($\mathcal{P}$-symmetry). SHG intensity shows a linear-in temperature increase across the altermagnetic transition, indicating that the $\mathcal{P}$-symmetry is preserved.
Therefore, the observed band splitting should be attributed to lifted Kramers{\textquoteright} degeneracy due to broken time-reversal symmetry.

Ideally, the observation of spin polarizations in the split bands by spin-resolved ARPES (SARPES) would be the most direct verification of altermagnetic state.
However, it is often challenging to measure spin polarization of time-reversal symmetry broken states by SARPES due to the formation of multiple domains which contribute to opposite signs of spin polarizations. With the beam spot from a He discharge lamp being in the $mm$-scale which is larger than the typical size of magnetic domains by orders of magnitude, one would usually obtain collective signals from multiple domains which average out to zero net polarization. Unlike ferromagnets where a large portion of domains can be aligned by applying external magnetic field~\cite{Osterwalder2006,Sumida2020,Hahn2021}, there is no universal way available at the moment to align antiferromagnetic or altermagnetic N\'eel vectors. Several methods including external magnetic field~\cite{Sapozhnik2018}, strain~\cite{Sapozhnik2017} and spin orbit torque~\cite{Wadley2016,Wadley:2018aa} are exploited; however, such methods are highly materials-specific and there is no general solution to aligning altermagnetic N\'eel vectors to yield macroscopic spin-polarized responses.

In the case of MnTe, a sixfold rotational symmetry in the $ab$-plane allows for 6 domains corresponding to N\'eel vector $\mathbf{L}\parallel[1\overline{1}00]$ and five equivalent directions. Such domains can be reduced to two antiparallel N\'eel vectors by external magnetic field~\cite{Kriegner2016}; however, one cannot further detwin the two antiparallel domains that are equal in energy. Therefore, alternative ways to align altermagnetic domains in MnTe are required to enable spin-resolved spectroscopic study. Using a ferromagnetic substrate has been suggested as a possible approach~\cite{Mazin2023}. High-quality MnTe film is an ideal platform to apply various switching methods, making it a promising workhorse material to explore the electronic structure and macroscopic properties of altermagnets on a single domain. 

In conclusion, by demonstrating momentum-resolved spectroscopic measurements and the spin split electronic structure in an altermagnetic MnTe for the first time, our study presented in this Letter lays firm groundwork for further studies on the novel magnetic ground state and potential future applications based on its unique properties.

This work was supported by the Institute for Basic Science in Korea (Grant No. IBS-R009-G2) and the National Research Foundation of Korea (NRF) grant funded by the Korea government (MSIT). (No. 2022R1A3B1077234).
C.-J. K. was supported by the NRF (Grant No. 2022R1C1C1008200). L. S acknowledges support from JGU TopDyn initiative.

%

\end{document}